\documentclass[traditabstract]{aa} % for the abstract without structuration 
\usepackage{graphicx}
\usepackage{natbib}
\bibpunct{(}{)}{;}{a}{}{,} % to follow the A&A style

\usepackage{txfonts}
\newcommand{\ha}{H$\alpha$~}
\newcommand{\msun}{M$_\odot$~}
\newcommand{\rsun}{R$_\odot$~}
\newcommand{\chil}{CI97}
\begin{document}
   \title{Photospheric and chromospheric activity in V405\,And}

   \subtitle{An M dwarf binary with components on the two sides of the full convection limit}

\author{K. Vida \inst{1}$^,$ \inst{2}
\and
K. Ol\'ah \inst{2}
\and
Zs. K\H{o}v\'ari \inst{2}
\and
H. Korhonen\inst{3}
\and
J. Bartus \inst{4}
\and
Zs. Hurta \inst{1}$^,$ \inst{2}
\and
K.  Posztob\'anyi \inst{5}
}

\institute{E\"otv\"os Lor\'and University, Department of Astronomy, H-1518 Budapest, PO Box 32, Hungary\\
\email{vidakris@elte.hu}
\and
Konkoly Observatory of the Hungarian Academy of Sciences, 1525 Budapest, PO Box 67, Hungary
\and
European Southern Observatory, Karl-Schwarzschild-Stra\ss e 2, 85748 Garching bei M\"unchen, Germany
\and
Astrophysical Institute Potsdam, An der Sternwarte 16, 14482 Potsdam, Germany
\and
AEKI, KFKI Atomic Energy Research Institute, Thermohydraulic Department, H–1525 Budapest 114, PO Box 49, Hungary
}

   %\date{Received September 15, 1996; accepted March 16, 1997}

% \abstract{}{}{}{}{} 
% 5 {} token are mandatory
 
  \abstract
   {We investigate the fast rotating ($P_{orb}=P_{rot}=0.465d$ ) active dwarf binary V405\,And (M0V+M5V) using  photometric $BV(RI)_C$ and optical spectroscopic data. The light variation is caused by the combined effect of spottedness and binarity with a small eclipse. From the available light and radial velocity curves we estimate the system parameters. Three flare events happened during the observations: two were found in the spectroscopic data and one was observed photometrically in $BV(RI)_C$ colours. An interesting eruptive phenomenon emerged from the photometric measurements which can be interpreted as a series of post-flare eruptions lasting for at least 3 orbits (rotations) of the system, originating from trans-equatorial magnetic loops, which connect the active regions in the two hemispheres. The two components of V405\,And have masses well over and below the theoretical limit of full convection. This rare property makes the binary an ideal target for observing and testing models for stellar dynamo action.}


   \maketitle
\sloppy
\section{Introduction}
Low mass stars are in the focus of studying stellar structure: these kind of objects make up the majority of the Galaxy, yet, because of their low luminosity, it is not easy to study them in detail.
Low mass stars with fast rotation show magnetic phenomena (spots, plages, flares, strong X-ray emission) due to some kind of magnetic dynamo in their interiors. Stars with radiative core and convection zone are thought to operate  $\alpha\Omega$ dynamos. The $\alpha$-effect creates poloidal field from toroidal one by helical turbulence in the convection zone, whereas $\Omega$-effect produce toroidal field from poloidal one by differential rotation. The place of the dynamo process is thought to be a thin layer between the core and convection zone, called tachocline, where the shear is the strongest. Below about $\sim$0.32\msun stars are thought to be fully convective, yet show all known magnetic phenomena as well. The source of the magnetic field in these low mass stars could be $\alpha^2$ dynamo generating strong, long-lasting, axisymmetric magnetic fields by turbulence, or by some other mechanism involving differential rotation. However, the mass limit (or the spectral limit of ~M3.5) of the full convection is not well established, since strong magnetic fields may shift this boundary towards lower masses (\citealt{2001ApJ...559..353M}, \citealt{2007A&A...472L..17C}). For low mass ($M < 0.8$) active stars a significant discrepancy is found between models and observations of stellar radii, i.e., theory predicts $\sim$10\% lower radii, while temperatures are underestimated by about 5\% \citep[see e.g.][]{2006Ap&SS.304...89R}.

An interesting approach is to study low mass stars above and below this hypothetical mass limit together in close binary systems. As to our knowledge, only two such systems are known. 2MASS J04463285+1901432, in NGC 1647 \citep{2006AJ....131..555H} has an orbital period of $0.619d$ and masses of 0.47 and 0.19M$_\odot$, however, its $V=19.34$ prevents more detailed investigation of this binary.

The binary studied in this work is an X-ray bright object, listed in the ROSAT All-Sky-Survey as RX J0222.4+4729 \citep{1996yCat.9011....0V}, later named as V405\,And \and classified as RS CVn type in the General Catalogue of Variable Stars \citep{2000IBVS.4870....1K,2009yCat....102025S}. The only detailed study of V405\,And is from \cite{1997A&A...326..228C} (\chil~ hereafter), who find that the binary has an orbital period of $P=0.465d$ showing a small (near grazing) eclipse, and present photometric observations in $B$ and $V$, radial velocity data and H$\alpha$ study. 
The components have spectral types M0V and M5V derived by \chil, both of them showing strong H$\alpha$ emission lines, of which the primary is modulated by the rotation. The original classification of the system is BY~Dra binary by the above authors. Note, that the definitions of the 'RS' and 'BY' types overlap in the GCVS, and for V405\,And both classifications seem to be valid.
 In \chil~orbital elements and estimates of the masses around 0.2 and 0.5M$_\odot$, and radii of 0.78 and 0.25--0.30\rsun are given as well, which were used as starting point for our analysis.

\section{Observations and data reduction}
\begin{figure}
 \centering
 \includegraphics[width=0.48\textwidth]{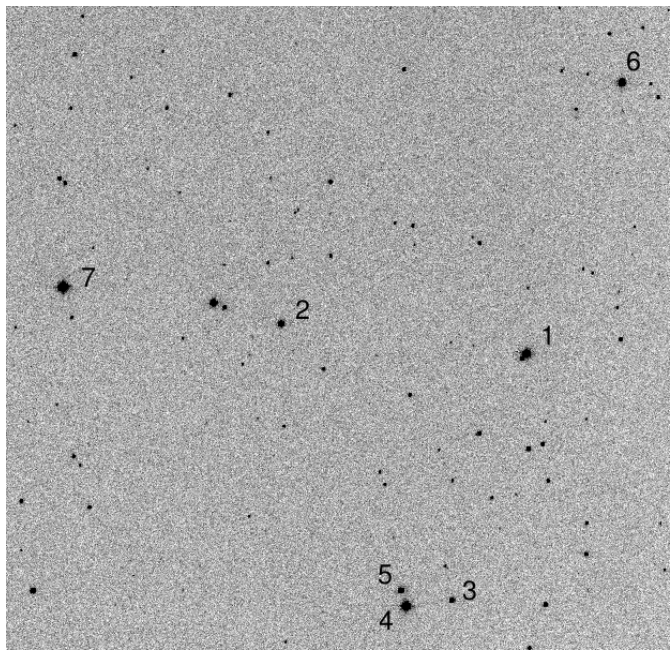}
 \caption{Finding chart for V405\,And from a $V$-band frame of the 1m RCC telescope. Numbering is the same as in \cite{1997A&A...326..228C} .}
 \label{fig:chart}
\end{figure}

Photometric observations were carried out using the 1m RCC telescope of Konkoly Observatory at Piszk\'estet\H{o} Mountain Station equipped with a Princeton Instruments $1300\times1300$ CCD and the 60cm telescope of Konkoly Observatory at Sv\'abhegy, Budapest with a Wright Instruments
$750\times1100$ CCD camera. 

Observations were collected during 28 nights: 13 nights with the 60cm telescope (February 8 -- October 1, 2007; 2454140--2454375 HJDs), and 18 nights with the RCC telescope (September 28, 2007 -- November 18, 2008; 2454372--2454789 HJDs), on three nights both telescopes were used simultaneously.

After correcting for atmospheric extinction and transforming the data to standard photometric system, we matched the two data series by fine-tuning the telescope constants within their errors.

A finding chart is plotted on Fig.~\ref{fig:chart}, V405\,And is marked as Star 1 (using the numbering of \chil). Two of the comparison stars (Star 3, Star 6) used by \chil~ turned out to be variable. Star 3 shows changes on a longer timescale of 20-30 days. Star 6,  showing a period of $\sim0.658d$, is possibly a W UMa type star. This effect would only cause small shifts, if any in \chil, since they used the average value of the five comparison stars.

In this work GSC 03298-00148 (Star 7) was used as a comparison star. Magnitudes of Star 7 were calculated  using stars 2,3,4 and 5 separately and the results were compared. The differences were 0.03 magnitudes at maximum. According to the results, we chose finally Star 2 for defining the magnitudes of the comparison star.

As mentioned in \chil, there is a star near V405\,And, at about 4.5 arc-seconds to the south-east, which is blended with the variable on most of the frames. We determined the magnitudes of the star near V405\,And using frames taken on 2007 September 29, October 01 and October 02. IRAF\footnote{
IRAF is distributed by the National Optical Astronomy Observatory, which
is operated by the Association of Universities for Research in Astronomy, Inc.,
under cooperative agreement with the National Science Foundation.
} 
PHOT routine of DAOPHOT package with small aperture, and NSTAR PSF-fitting routine were used. There was a systematic difference (order of 0.01 magnitudes) between aperture and PSF photometry (aperture photometry gave brighter magnitudes). Since the two stars are separated only by 12.6 pixels, V405\,And extends into the aperture of the fainter star, so the result of the PSF photometry was accepted. The magnitudes of the faint star are
$16.52\pm0.06$,
$14.92\pm0.05$ (15.1 in \chil),
$13.75\pm0.03$ and
$12.13\pm0.03$
in $B,V,R_C$ and $I_C$ passbands, respectively. To test the correctness of these values, we compared PSF-magnitudes of V405\,And with magnitudes from aperture photometry. We set an aperture large enough to contain both V405\,And and the faint star, then subtracted the intensity of the faint star from the result. These magnitudes from aperture photometry and PSF-photometry match reassuringly within a few hundredths of magnitude.

\begin{figure}
 \centering
\includegraphics[width=0.48\textwidth]{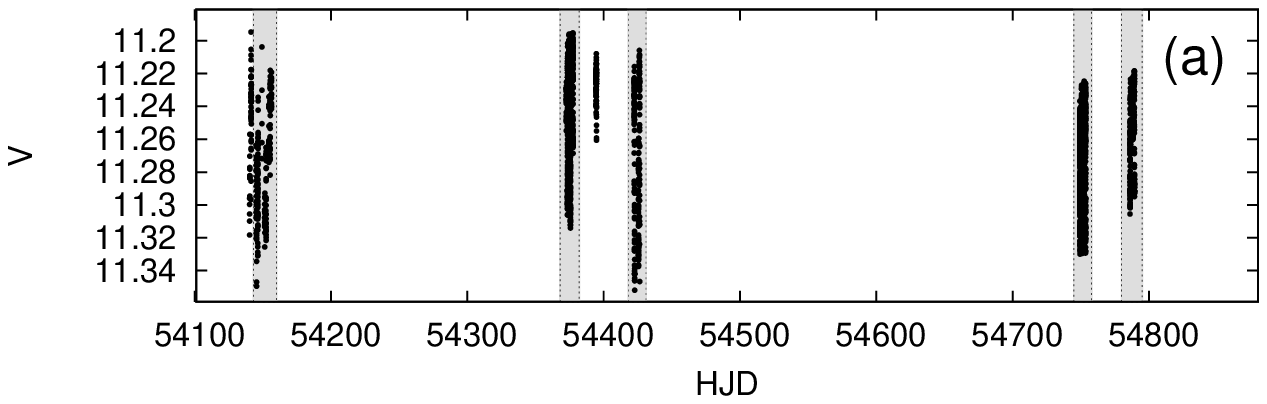}
 \includegraphics[width=0.4725\textwidth]{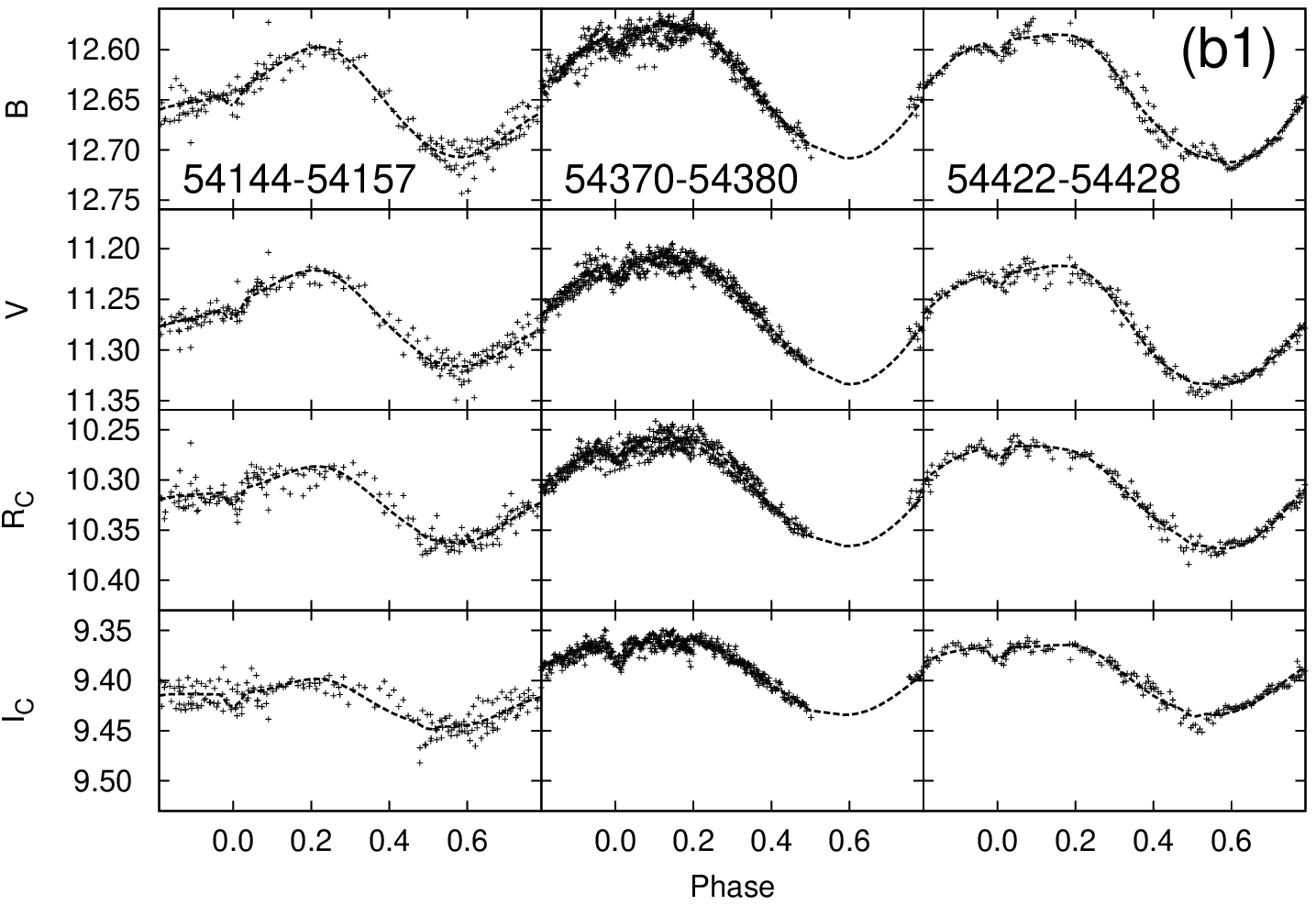}
 \includegraphics[width=0.336\textwidth]{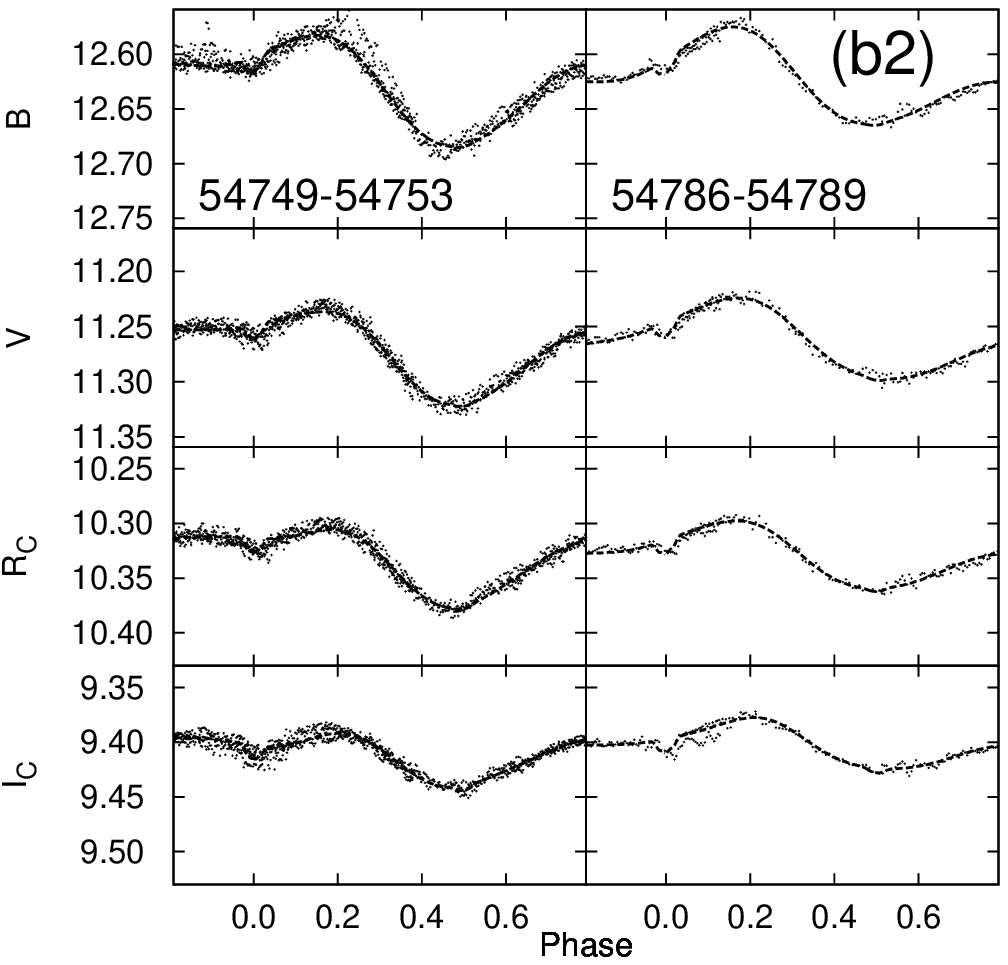}
 \label{fig:slices}
 \caption{a) Light curve in $V$ passband. The scale is smaller than the actual changes of the star caused by a flare at the beginning of the observation. b1,b2) Four-colour light curves of V405\,And, and the fitted light curve solution containing both binarity and spottedness. The phased light curves contain observations from Julian date intervals shown on the plots.}
\end{figure}

Differential aperture photometry of V405\,And was done using DAOPHOT routine. The aperture was chosen to contain the faint star on all frames, since the two stars were not resolved on most frames. The intensities of this star were subtracted from the light curve after transforming to standard Johnson--Cousins photometric system. The resulting light curve is plotted in Fig.~\ref{fig:slices}.

Interstellar extinction was checked for V405\,And using NASA/IPAC Extragalactic Database (NED), giving a value of $E(B-V)=0.182$ mag for total extinction. According to \chil, the distance of V405\,And is 25--35 pc, so the effect of interstellar extinction would be only a few thousandth magnitude. This value is much smaller than the errors of the photometry ($\sim$0.01--0.02mag), so it can be neglected.

Altogether 97 high-resolution spectra was retrieved from the ELODIE archive at Observatoire de Haute-Provence (OHP) \citep{2004PASP..116..693M}. In 1994 38 spectra were observed: 27 between October 18--23, and 11 between December 12--16. Another 59 unpublished spectra were obtained between October 25--30, 1998. The spectra from the archive cover the wavelength range of 4000--6800\AA~with a resolution  ($\lambda/\Delta\lambda$) of 42000.

%%%%%%%%%%%%%%%%%%%%%%%%%%%%%%%%%%%%%%%%%%%%%%%%%%%%%%%%%%%%%%%%%%%%%%%%%%%%%%%%%%%%%%%%%%%%%%

\section{Analysis}
\begin{table}
\caption{Average magnitudes of stars on V405\,And field.}

\begin{center}
\begin{tabular}{lcccc}
\hline
Star No.$^*$&	$V$&	$B-V$&	$V-R_C$&	$V-I_C$\\
\hline
1 (V405\,And)&     11.25&	1.36&	  0.94&	    1.86\\ %with all data

&    \textit{11.10}$^{**}$&     \textit{1.44}&     \textit{0.89} &     \textit{1.74}
\medskip \\
2&     12.80&     0.43&     0.28&     0.55\\
 &     \textit{12.80}&     \textit{0.43}&     \textit{0.28}&     \textit{0.55}
\medskip \\

3&     13.60&     0.43&     0.28&     0.57\\
 &     \textit{13.60}&     \textit{0.41}&     \textit{0.28}&     \textit{0.54}
\medskip \\

4&     11.05&     0.44&     0.28&     0.54\\
&     \textit{11.05}&     \textit{0.43}&     \textit{0.28}&     \textit{0.54}
\medskip \\

5&     12.99&     0.89\\
&     \textit{12.96}&     \textit{0.94}
\medskip \\

6&     11.67&     0.68&     0.45&     0.83\\
&     \textit{11.65}&     \textit{0.72}&     \textit{0.42}&     \textit{0.79}
\medskip \\

7&     10.67&     0.76&     0.47&     0.87\\
\end{tabular}\\
\smallskip
\end{center}
$^*$ The star numbers are the same as in \chil. Star 7 is not in the field of view in \chil.
\\
$^{**}$ Values in italics are from \chil.
\label{tab:ChIl-RCC}
\end{table}

Results in Table \ref{tab:ChIl-RCC} show that while the average magnitudes of our measurements agree quite well with those of \chil~ in case of Stars 2 to 6, values for V405\,And seem to differ significantly. V405\,And was 0.15 magnitude fainter during our observations in $V$ band than in 1995. This, and the decrease of $V-R_C$ and $V-I_C$ indices indicate higher spottedness. In the same time, the bluer $B-V$ show the presence of more faculae. Thus, we conclude that the activity is enhanced compared to the 1995 level.

\subsection{Spots: photometric modelling}

For finding light curve solution we adopted the method described by \cite{2008A&A...485..233S}. The method uses an iterative modelling for the binarity and spottedness separately. The binary model was computed using PHOEBE \citep{2005ApJ...628..426P}  simultaneously for our four-colour measurements and the radial velocity curve from \chil. The binary model light curve is removed from the observed one using the following equation:
\begin{equation}
 F_\mathrm{spot}=F_\mathrm{obs}-F_\mathrm{model}+F_\mathrm{unsp},
\end{equation}
where $F_\mathrm{spot}$ is the flux of the primary changing due to spottedness, $F_\mathrm{obs}$ is the observed light curve, $F_\mathrm{model}$ is the binary model constructed with PHOEBE, and $F_\mathrm{unsp}$ is the unspotted flux of the system (for model parameters see Table \ref{tab:params}). $F_\mathrm{spot}$ is a light curve which contains only the changes caused by the spottedness, and is free from the effects of binarity (eclipse and changes by the distorted shape of the stars). The true spotted light curve was modelled with {\sc SpotModeL} \citep{2003AN....324..202R}, which uses the analytic approach with circular spots of \cite{1977Ap&SS..48..207B}. The circumpolar part of the spotted region accounts for the long-term, overall brightness change, whereas spotted parts outside this region cause the rotational modulation. The latitude of the spots is, at least partly, determined by the ratio of buoyancy and Coriolis-force. On fast rotating stars the Coriolis-force is dominant which moves the spots to high latitudes \citep{1987ApJ...316..788C}. 
While spot models can determine spot longitudes quite reliably, photometric observations have very low information content on spot latitudes  \citep[see e.g.][]{1997A&A...323..801K}.
Assuming circular spots on the surface is an approximation with small number of parameters. Sunspots and active nests on the Sun give examples of circular active regions. Note, that as 'spots' we mean regions with dark spots and bright faculae, the (average) temperature of the region is made up from the cool and hot parts, relative to the photosphere.

\begin{figure}
 \centering
 \includegraphics[width=0.35\textwidth]{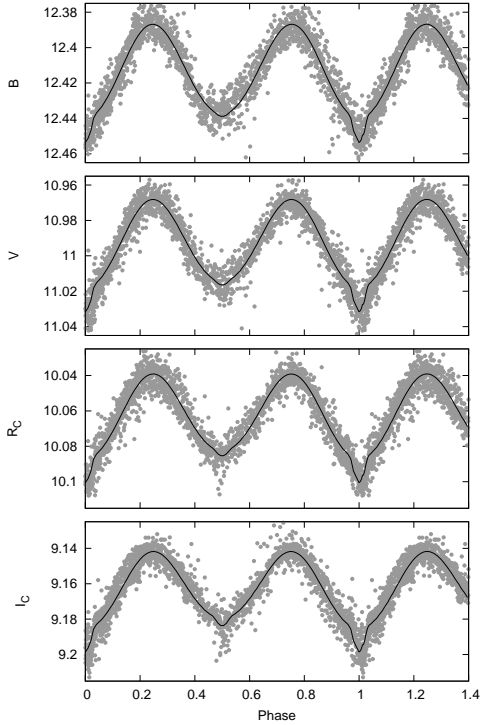}
 \caption{Phased $BV(RI)_C$ light curves from 2007 after subtracting the spot model from the observations. This dataset was used for determining the binary system parameters with PHOEBE.}
 \label{fig:phoebe-model}
\end{figure}

We assumed two spotted regions on V405\,And with average $T_\mathrm{spot}=$3300K on the high-latitude regions of the primary. 
Tests of \cite{1997A&A...323..801K} showed that analytic models using two circular spots describe the observed spotted light curves satisfactorily within the precision of the photometric observations.
Spot temperature was chosen by modelling a $V-I_C$ light curve, and the result was accepted for all passbands. For finding the best fitting spotted light curve, the size and position of the spots were searched. 
 During our observations the star was more than 0.1 magnitudes fainter due to higher spottedness than the faintest observed value in \chil, thus we used the brightest part of the $V$ and $B-V$ light curves of \chil~ as the (faintest possible) unspotted $B$ and $V$ magnitudes. Unspotted $R_C$ and $I_C$ magnitudes were calculated from the color indices.
The resulting spotted curve was then removed from the observed one, and these steps were repeated until we got a satisfactory fit for the spots and binarity (Fig.~\ref{fig:phoebe-model}) separately, and for the observed light curve  itself (Fig.~\ref{fig:slices}b).

\begin{table}
\caption{Binary parameters from the system modelling and spot model parameters.}
\begin{center}
\begin{tabular}{rcl}
$P_{orb}=P_{rot}$ &=& $0.46543d$ $^*$ \\
a &=& 2.246$R_\odot$  $^{**}$\\
$q$&=& $0.42\pm 0.01$\\
$i$&=& $66.5\pm 1^\circ$\\
$M_1$ &=& $0.49 M_\odot$\\
$M_2$ &=& $0.21 M_\odot$\\
$T_1$ &=& $4050 \pm 200$ K\\
$T_2$ &=& $3000 \pm 300$ K\\
$R_1$ &=& $0.78\pm0.02 R_\odot$\\
$R_2$ &=& $0.24\pm0.04 R_\odot$\\
$v\sin i$ &=& 85 km/s$^*$\\
$e$ &=& $0.0^*$ \\
\end{tabular} 
\begin{tabular}{rcl}

$B_\mathrm{unsp}$ &=& 12.42\\
$V_\mathrm{unsp}$ &=& 11.00\\
$R_\mathrm{unsp}$ &=& 10.07\\
$I_\mathrm{unsp}$ &=&  9.17\\
$T_\mathrm{spot}$ &=&  3300K\\
&&\\
&&\\
&&\\
&&\\
&&\\
&&\\
&&\\
\end{tabular}
\end{center}
\smallskip 
$^*$ adopted from \chil\\
$^{**}$  a$\sin i$ adopted from \chil
\label{tab:params}
\end{table}

Using this modelling method it is possible to determine the physical parameters of the binary (see Table \ref{tab:params}). For initial values of surface temperatures we used the tables of \cite{1996ApJ...469..355F} and \cite{2003AJ....126..778V}. The shape of the light curve is not very sensitive to the changes in the temperature, thus the fitting of these values is quite uncertain. Our result for the mass ratio is within $1\sigma$ from the value that was spectroscopically determined in \chil~ ($q=0.38\pm0.04$).

The spot configuration is relatively stable throughout the observations, positions and sizes of the spots change only slightly. 
According to the model, one spot is situated on the northern hemisphere, and another one on the southern hemisphere about 150$^\circ$ from each other in longitude.
From the modelling results the longitudes of spots (i.e., phases) are accurate within a few degrees, but latitude results are just approximations. Using the supposed $T_\mathrm{spot}$ we find that about 25\% of the stellar surface is covered with active regions. The stability of circular spot models are discussed in details in \cite{1997A&A...323..801K}. With the help of this simple spot model we could satisfactorily separate the light variation originating from binarity and activity.

\subsection{Hot regions in the chromosphere and photosphere}
\label{sec:chromosphere}
\begin{figure}
 \centering
 \includegraphics[width=0.5\textwidth,bb=50 50 410 302]{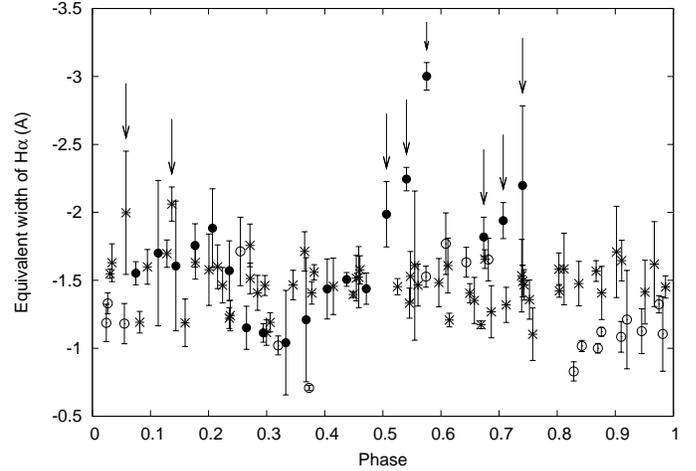}
 % eq_width.eps: 0x0 pixel, 300dpi, 0.00x0.00 cm, bb=50 50 410 302
 \caption{Equivalent widths of the \ha line. Data obtained on October 18, 1994 are plotted with filled points, empty points show measurements from October 21/23 and December of 1994, spectra from 1998 are plotted with crosses. Arrows show times of flares in 1994 and 1998 (see Section \ref{sec:flares}).}
 \label{fig:eq_width}
\end{figure}

\begin{figure}
 \centering
 \includegraphics[width=0.38\textwidth,angle=-90]{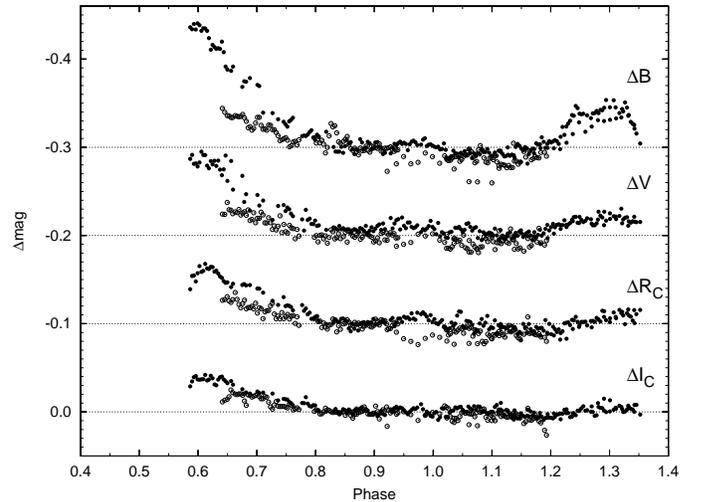}
 \caption{Light curve observed on September 29--30, 2007 at Piszk\'estet\H{o}. The binary and spot model is removed from the observed light curves. Dashed lines show zero level for each passband. Data from the 29th and 30th September are plotted with empty and full circles, respectively.}
 \label{fig:slowflare}
\end{figure}

\begin{figure}
 \centering
 \includegraphics[width=0.3\textwidth]{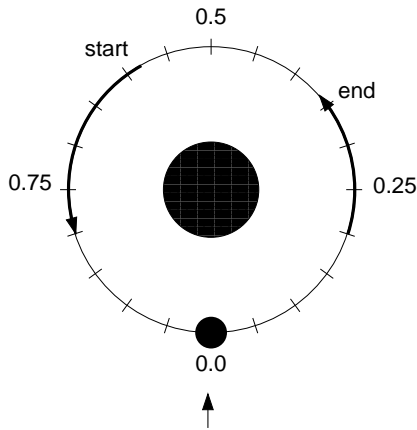}
 \caption{Schematic plot of the eruption times and visibility on 29--30 Sept, 2007. Arrow at the bottom shows the line of sight. The figure is plotted with appropriate scale.}
 \label{fig:slowflare_plot}
\end{figure}

The chromospheric activity of V405\,And was investigated using \ha line profiles.
Equivalent widths (Fig.~\ref{fig:eq_width}) were measured by fitting Gaussian profiles using SPLOT task in IRAF, after the continuum was fitted by polynomials and removed. The equivalent widths vary between 1--2\AA~showing a small rotational modulation. This variation seems to be fairly regular during the observations, except for the time of the flares in October 1994 and 1998 (see Section \ref{sec:flares}). The observed variation indicates permanent plage structure in the stellar atmosphere.

In September 29--30, 2007 an interesting phenomenon was observed (see $BV(RI)_C$ light curves in Fig.~\ref{fig:slowflare}). During two consecutive nights, at the beginning of the observations the binary showed excess brightness in all bandpasses which gradually decreased to the unspotted flux. After about one third rotation the brightening started again (the system configuration during the event is plotted on Fig.~\ref{fig:slowflare_plot}). The excess brightness was stronger during the second night (between the two sets of observations was one full rotation). The ratios of the peak intensities in $BV(RI)_C$ filters at the beginning of the observations differ from that produced by a usual flare event at or near its peak intensity.

Assuming that the brightening is caused by hot regions on the star  we can make a model of the surface using {\sc SpotModeL}. Results show that  two hot areas  separated in longitude by $\sim$90--120$^\circ$ describe the observed changes well. Making use of two passbands,
from $B-V$ and $V-I_C$ indices we get hot spot temperatures of $5130\pm270$K and $5680\pm300$K, respectively, which means these regions are at least 1000K hotter than the surrounding area. At maximum of the brightness $B-V=1.31$ yields 4300K for the temperature \citep{1996ApJ...469..355F}, meaning that the whole visible stellar surface seemed hotter by $\sim$150 K, i.e.; about 10--12\% of the visible surface of the star is covered with hot regions.

An explanation of the phenomenon could be a long-lasting (at least 1.5 days, or 3 rotations) flare event on V405\,And. 
During the first night's observations we see only the decreasing part of a giant flare that might have occurred during daytime, before the observations started. At the beginning of the second night we observe a peak which could then be a post-flare eruption. Two smaller eruptions happened later that night, the first was small, the second was more powerful and started to decrease still before the end of the night.

A similar long-lasting event has been observed on AU\,Mic. This event was described and modelled in \cite{1999ApJ...510..986K}. The authors,  comparing their results with observed solar phenomena, supposed  long-lasting giant post-flare loops with a size of 1--2 stellar radii, and energy balance through continuing reconnections.

\begin{figure}
 \centering
 \includegraphics[width=0.38\textwidth,angle=-90]{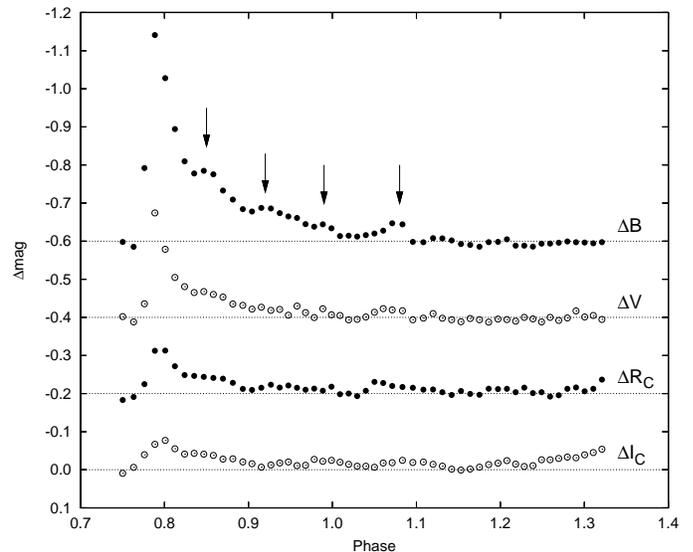}
 \caption{Flare observed on February 9, 2007 in Budapest. The binary and spot model is removed from the observed light curves. Light curves from up to down are made using $B, V, R_C, I_C$ filters. Dashed lines show zero level for each passband (an arbitrary shift was applied for better visibility). Arrows show the times of the post-flare events.}
 \label{fig:bigflare}
\end{figure}

Transequatorial magnetic loops are common features on the Sun. \cite{2003ApJ...598L..59H} studied the energy mechanism of such a solar loop system and found brightenings and flare-like events during 2 days of observations. On V405\,And from photometric observations we suggest 1-1 dominant active region in both hemispheres of the primary separated by $\sim$150$^\circ$ in longitude. We could explain the observed phenomenon on September 29--30, 2007 with a transequatiorial loop system which causes the brightenings and the flare-like events. As the star rotates, the emitting part of the loop system eventually turns in and out of view. This scenario would explain the turn-backs of the light curve seen, e.g., in $B$ band at 0.3 phase (Fig. \ref{fig:slowflare}), and also the intensity ratios in different filters, which are comparable to  the post-flare event of Fig.~\ref{fig:bigflare}.

\subsection{Flares}

\label{sec:flares}

\begin{figure*}
 \centering

\includegraphics[width=0.33\textwidth]{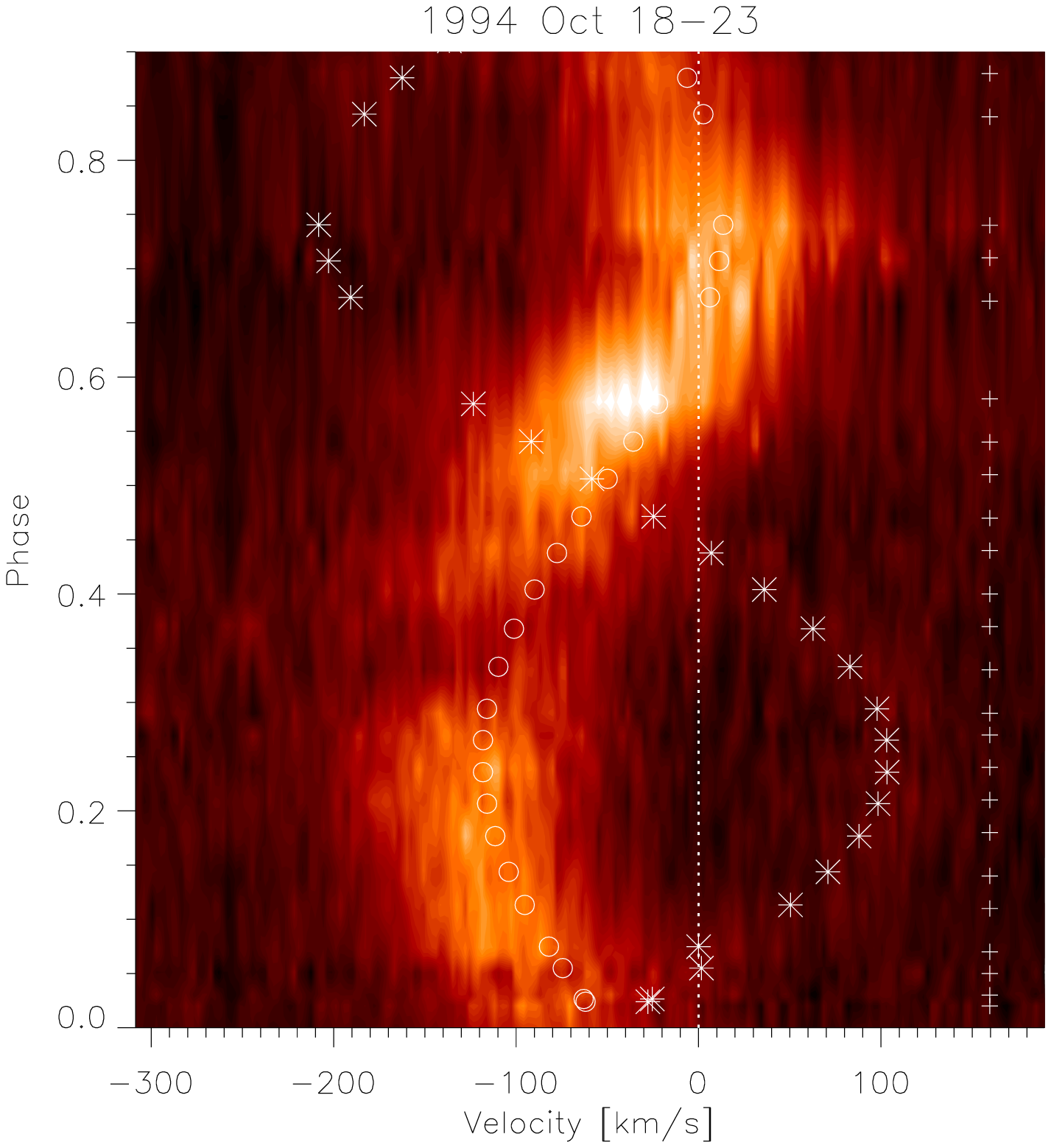}
\includegraphics[width=0.33\textwidth]{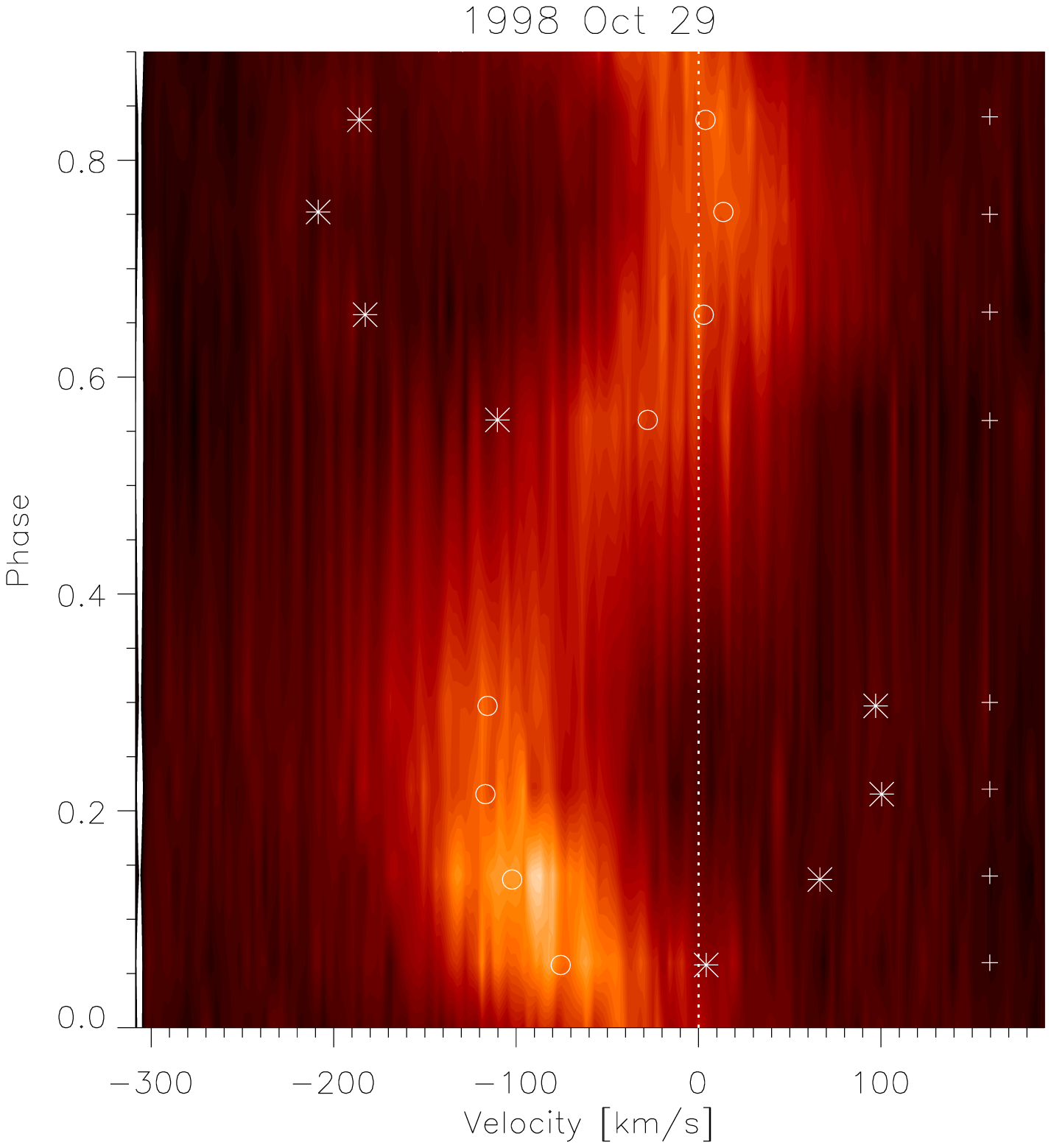}
\includegraphics[width=0.33\textwidth]{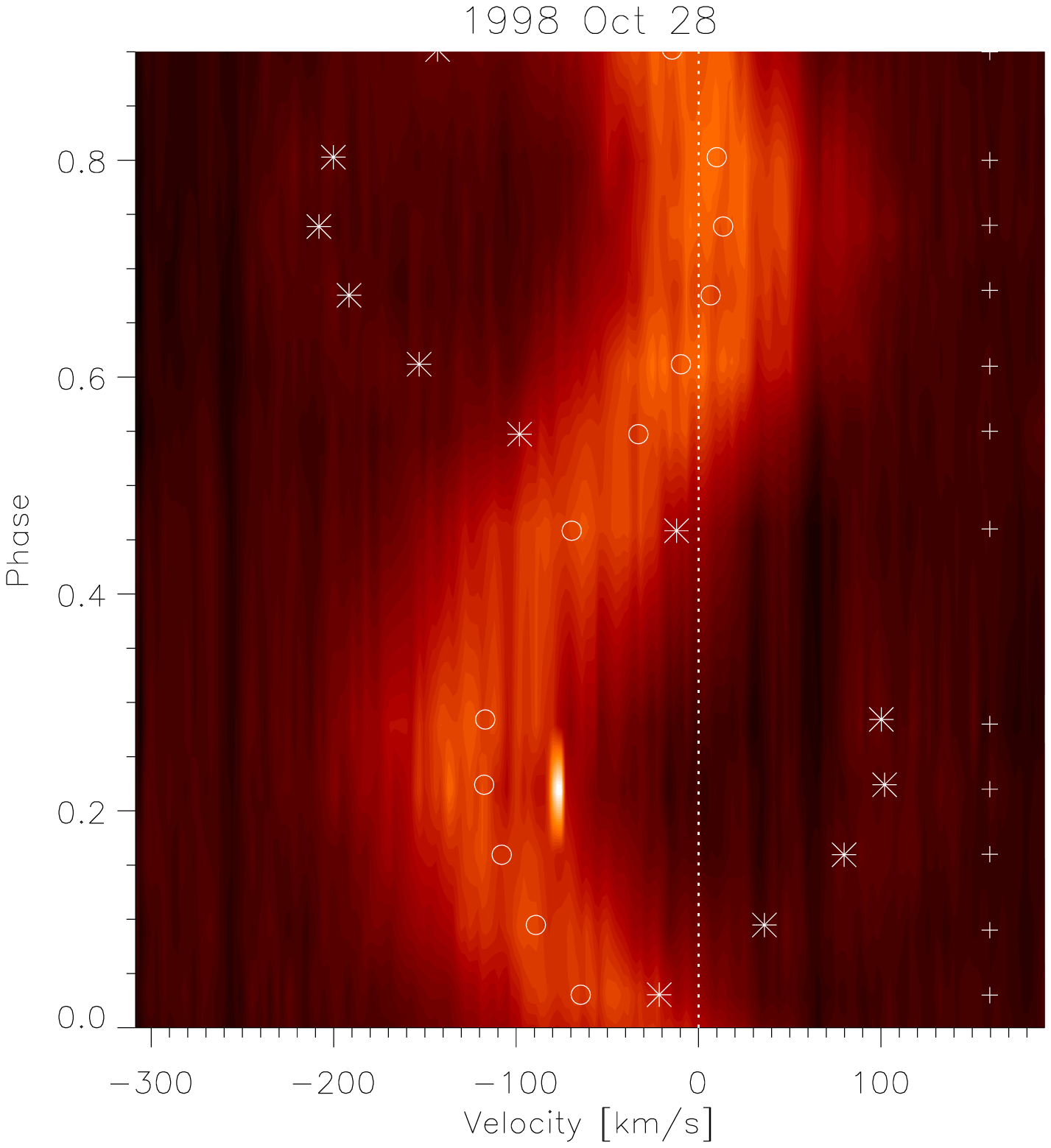}

 \caption{Dynamic \ha spectra from October 18 and 23, 1994, October 29 and 28, 1998 (from left to right). Crosses on the right show the time of observations, circles on the plot and stars mark the radial velocity curve of the primary and secondary, respectively. The first two images on the left show two flares, the third one is from a quiescent state of the binary. All plots show phases between 0.1--0.9. The left plot shows observations from October 18, 1994 except the last two points from October 23, 1994 after phase 0.75.}
 \label{fig:halpha}
\end{figure*}

\begin{figure*}
 \centering
 \includegraphics[width=0.33\textwidth]{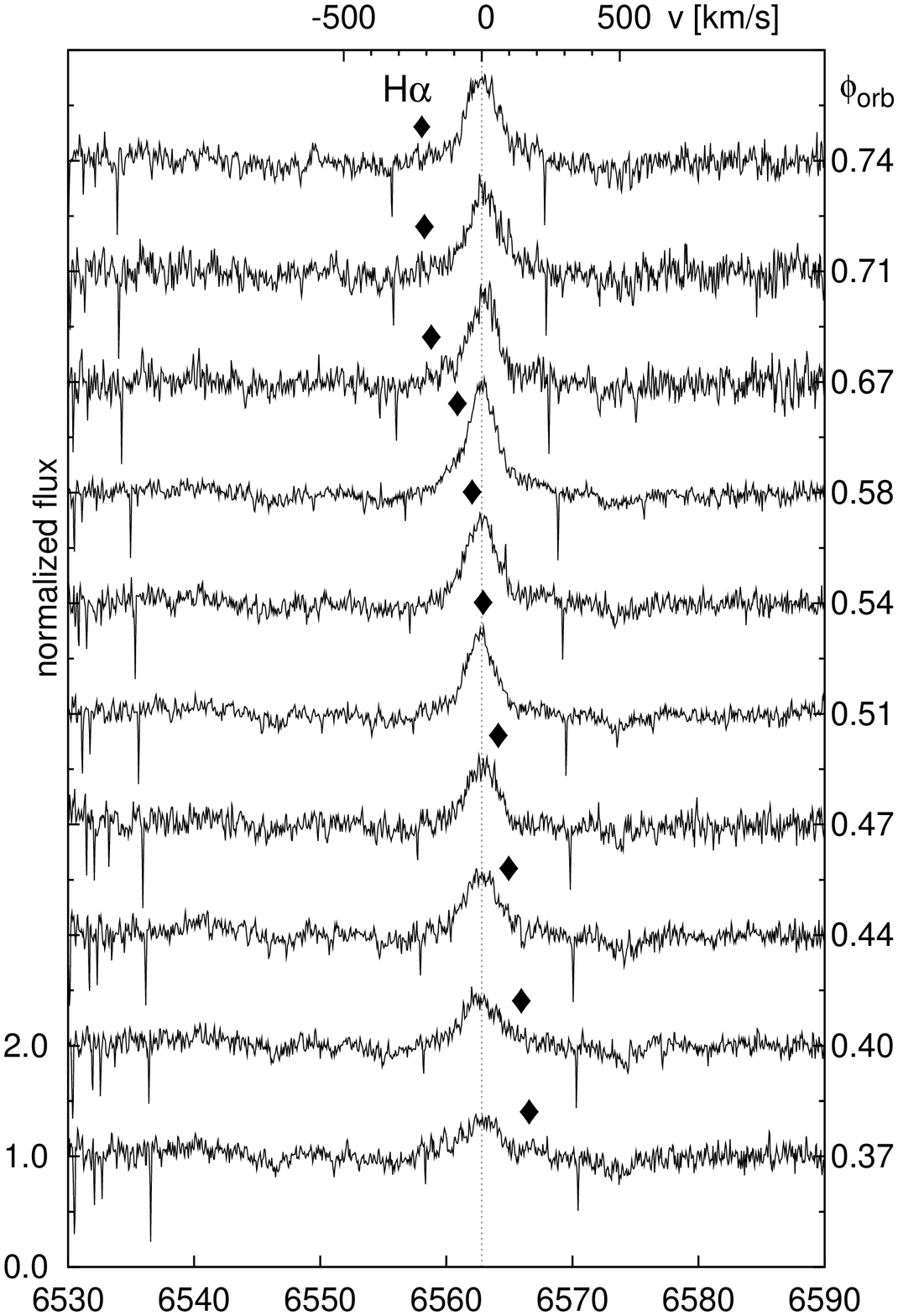}
 \includegraphics[width=0.33\textwidth]{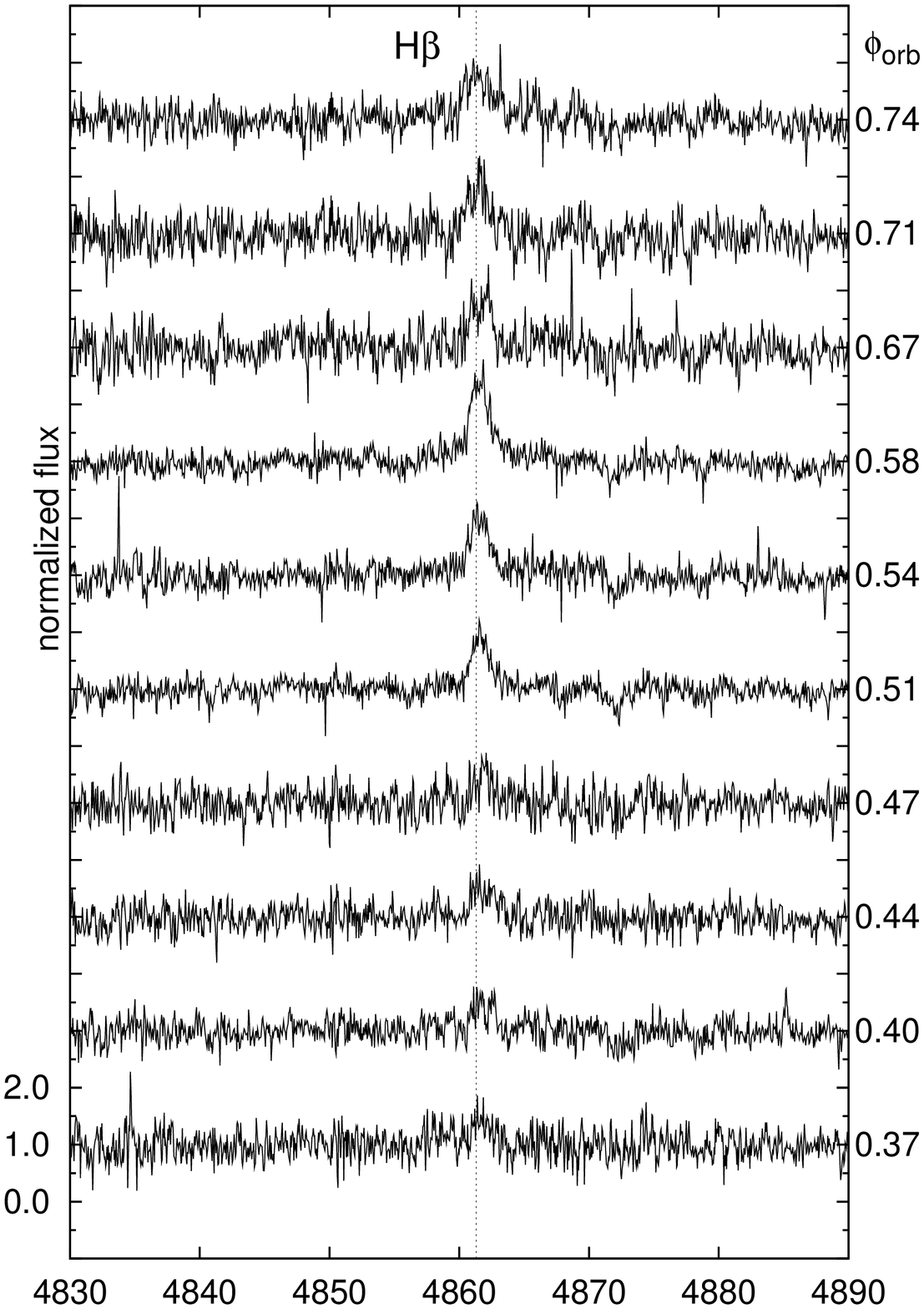}
 \includegraphics[width=0.33\textwidth]{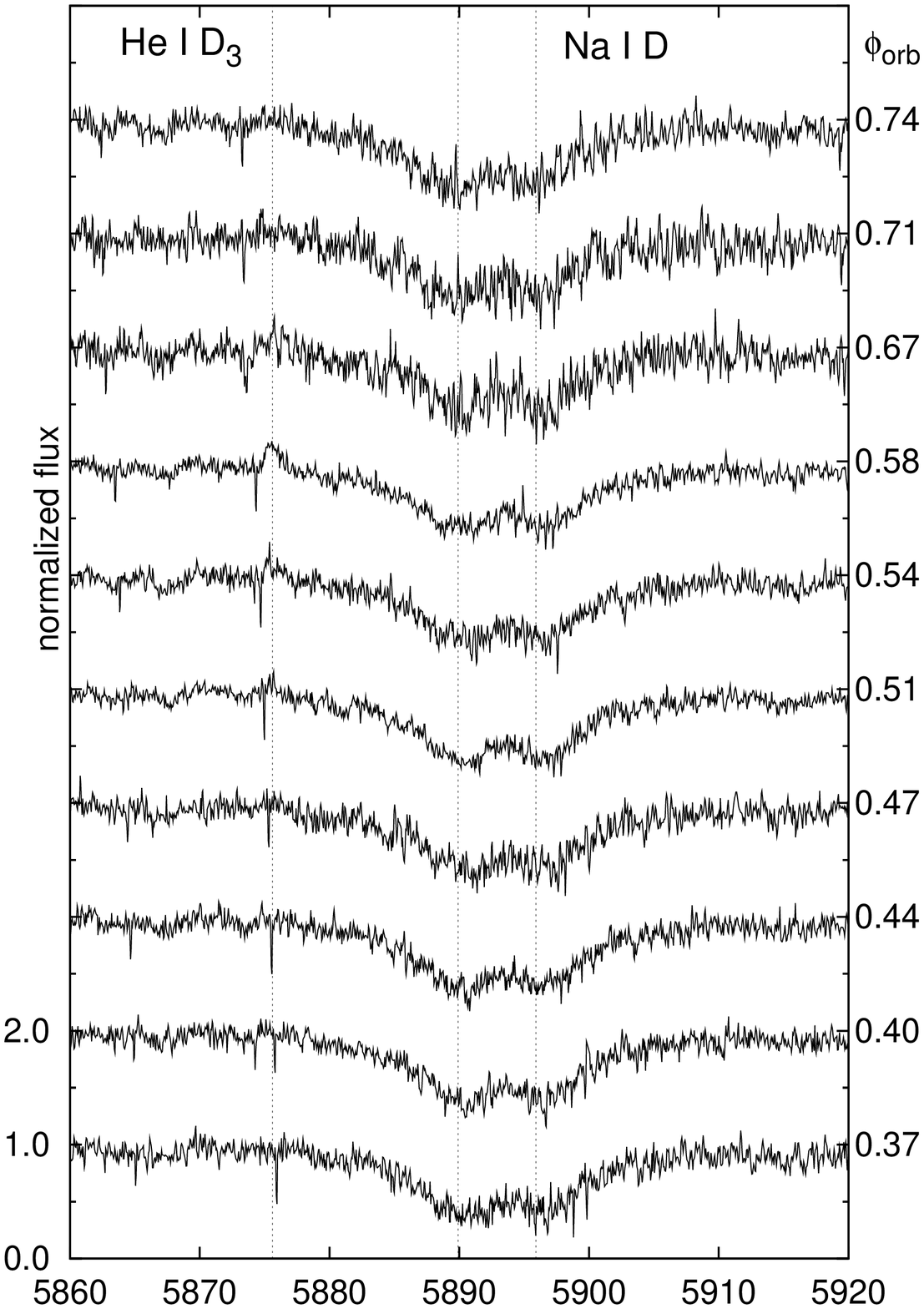}
 \caption{H$\alpha$, H$\beta$ and Na D lines around the time of the flare event in October 18, 1994 between phases 0.37--0.74. The flare occurred between phases $\sim$0.5--0.7 (see scale on the right). Note that the scale of the H$\beta$ plot is different because of the higher noise in the blue part of the spectrum. Diamonds on the \ha spectra show the place of the secondary component calculated from the radial velocity curve.}
 \label{fig:fler1994_spect}
\end{figure*}

\onlfig{10}{
\begin{figure*}
 \centering
 \includegraphics{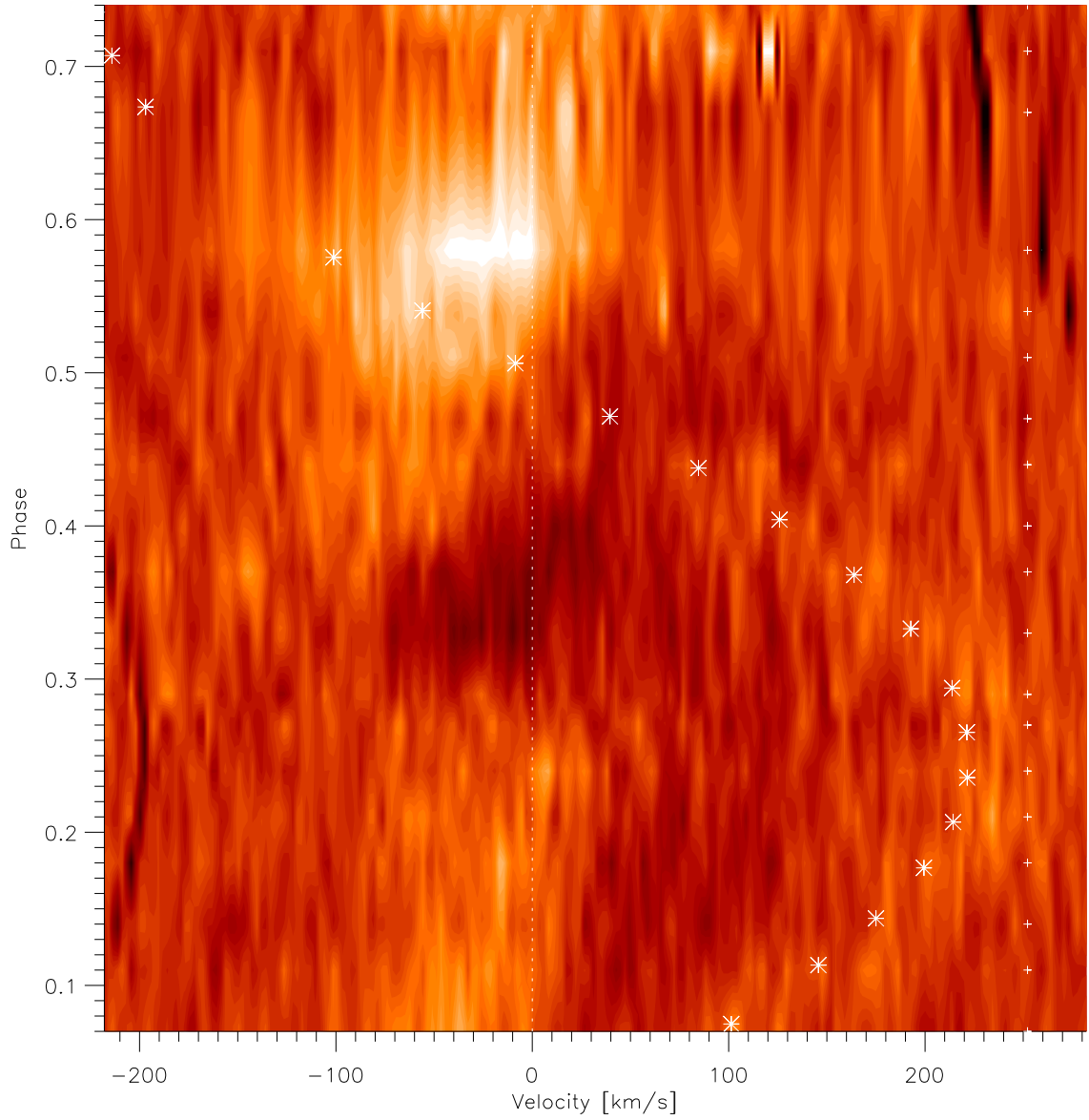}
 % 19941018dynamic.eps: 0x0 pixel, 300dpi, 0.00x0.00 cm, bb=0 0 453 453
 \caption{Dynamic \ha spectrum with the flare from October 18, 1994 showing the excess emission compared to the average of all spectra. The spectra are shifted with their actual radial velocity.}
 \label{fig_online:dynamic}
\end{figure*}
}

\onlfig{11}{
\begin{figure*}
 \centering
 \includegraphics[width=0.33\textwidth]{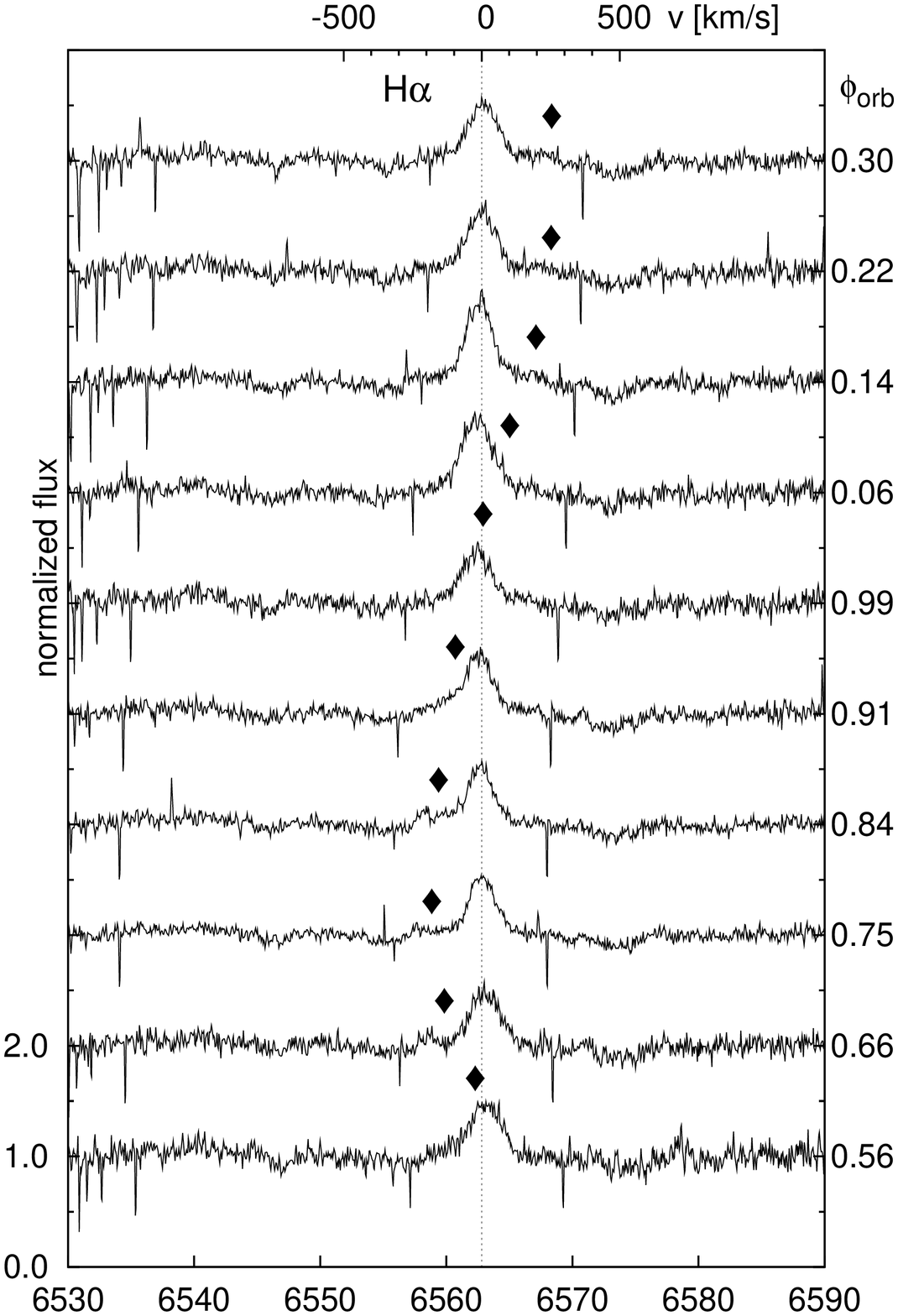}
 \includegraphics[width=0.33\textwidth]{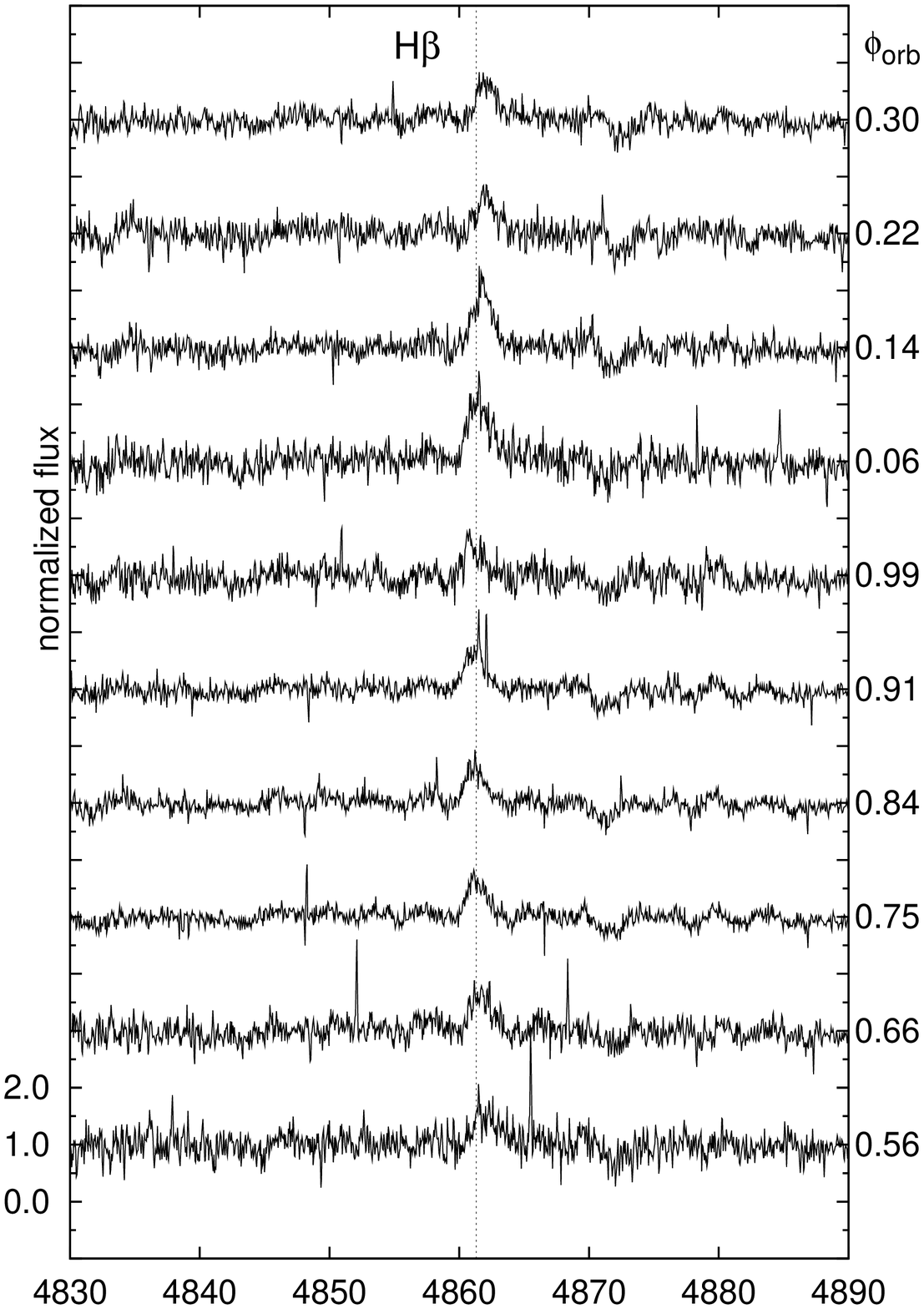}
 \includegraphics[width=0.33\textwidth]{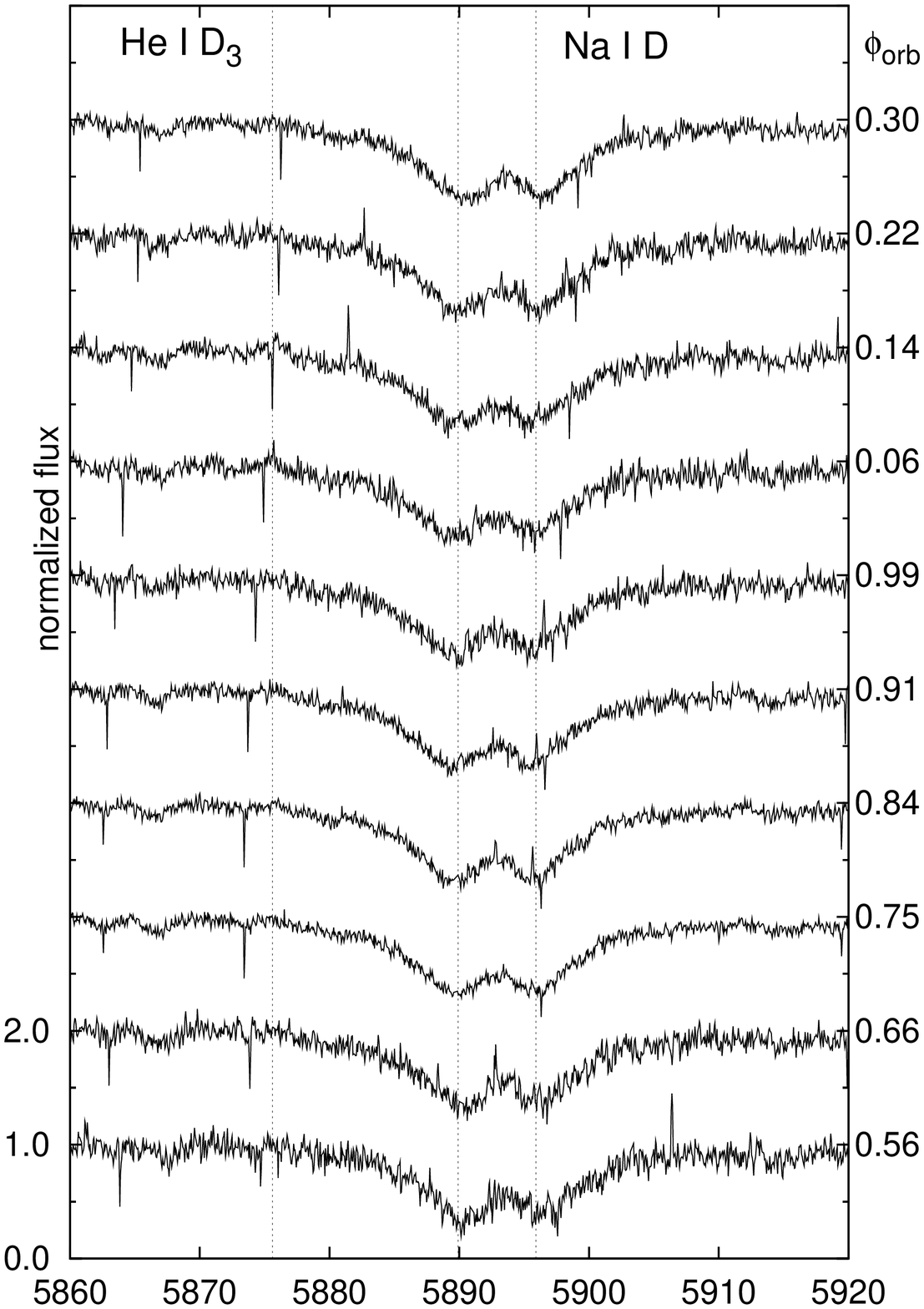}

 \caption{H$\alpha$, H$\beta$ and Na D lines around the time of the flare event in October 29, 1998 between phases 0.56--0.30. The flare occurred between phases $\sim$0.05--0.15 (see scale on the right). Scales and notation is the same as in Fig.~\ref{fig:fler1994_spect}. }
 \label{fig_online:fler1998_spect}
\end{figure*}
}

\onlfig{12}{
\begin{figure*}
 \centering
 \includegraphics[width=0.33\textwidth]{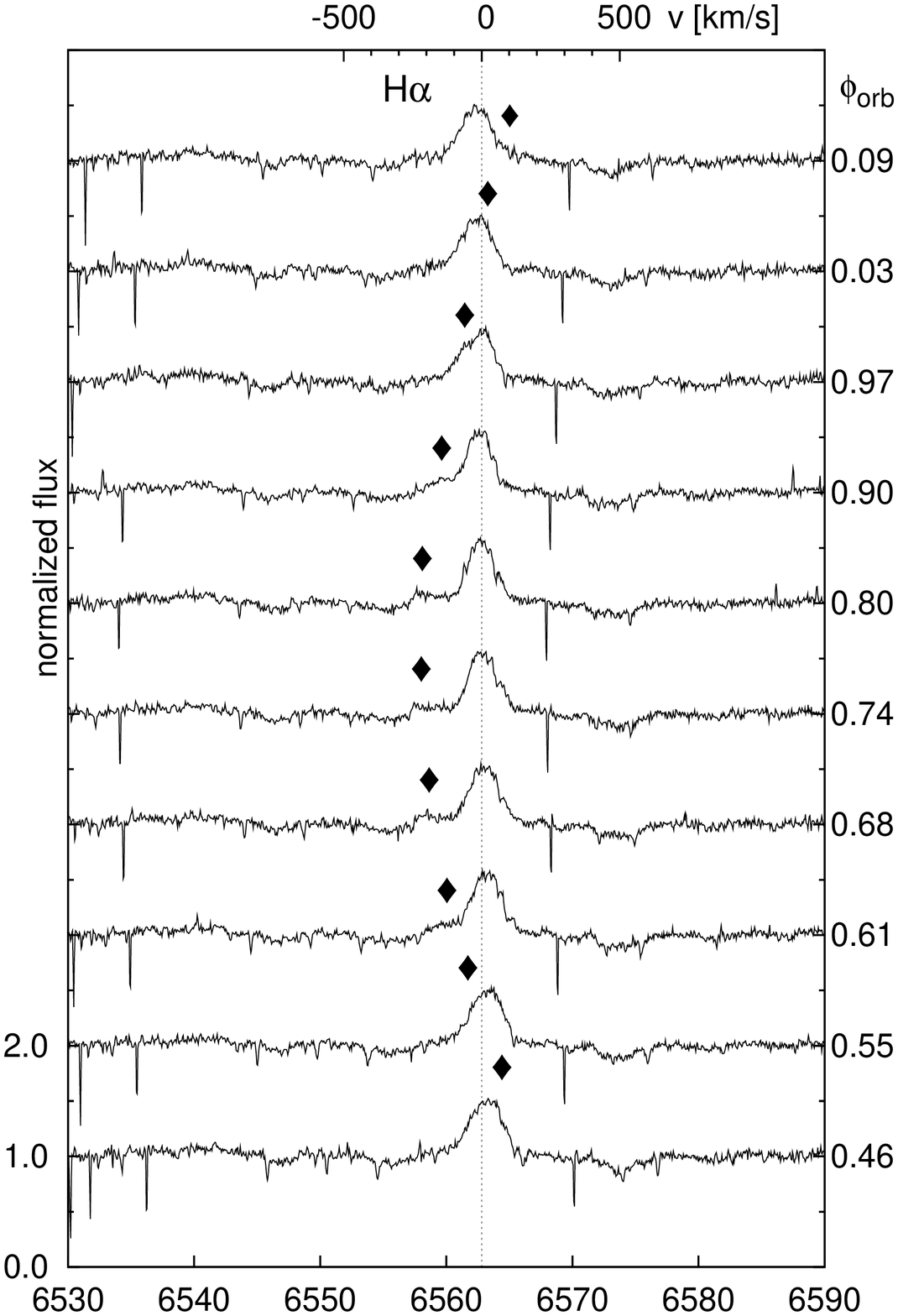}
 \includegraphics[width=0.33\textwidth]{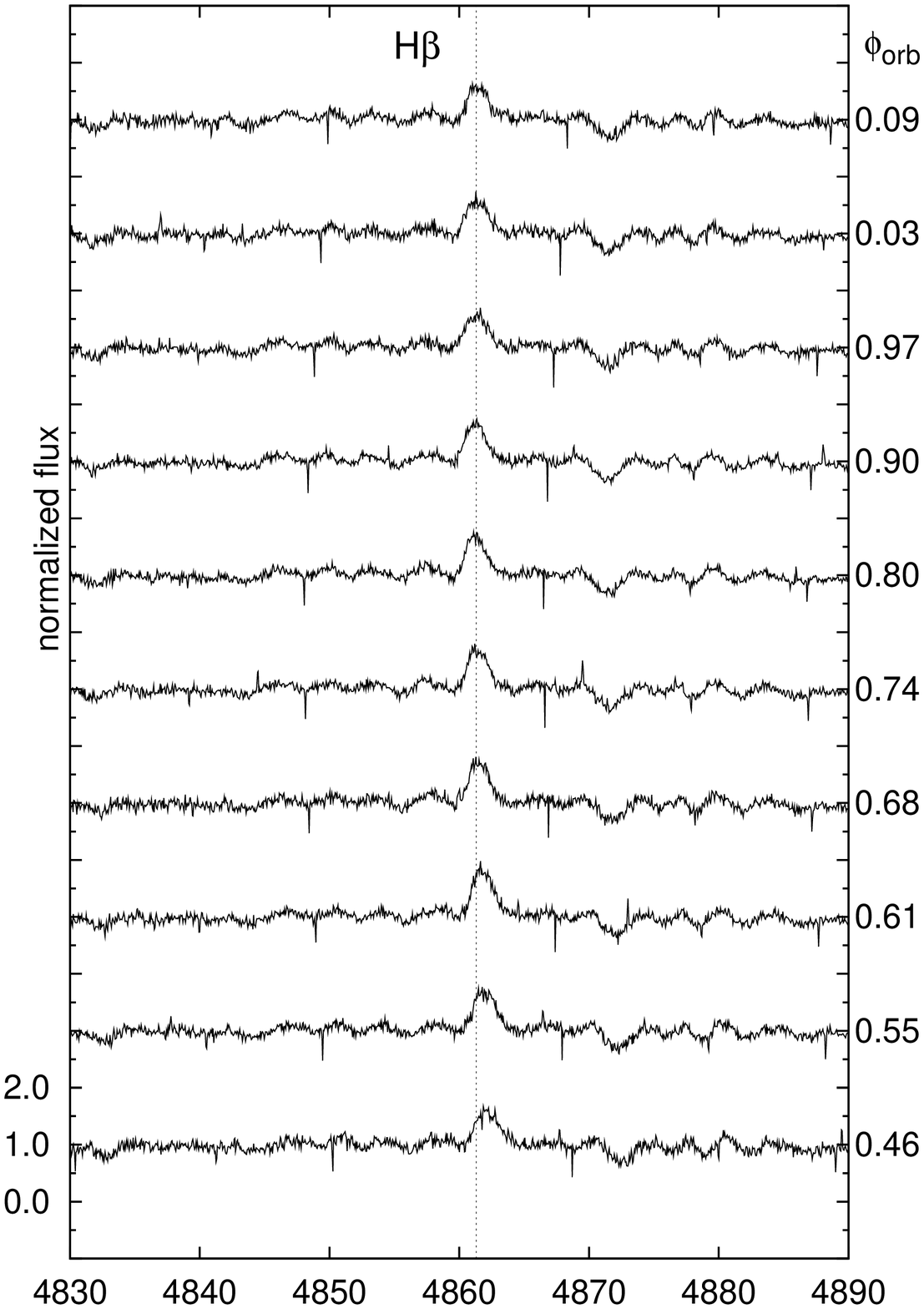}
 \includegraphics[width=0.33\textwidth]{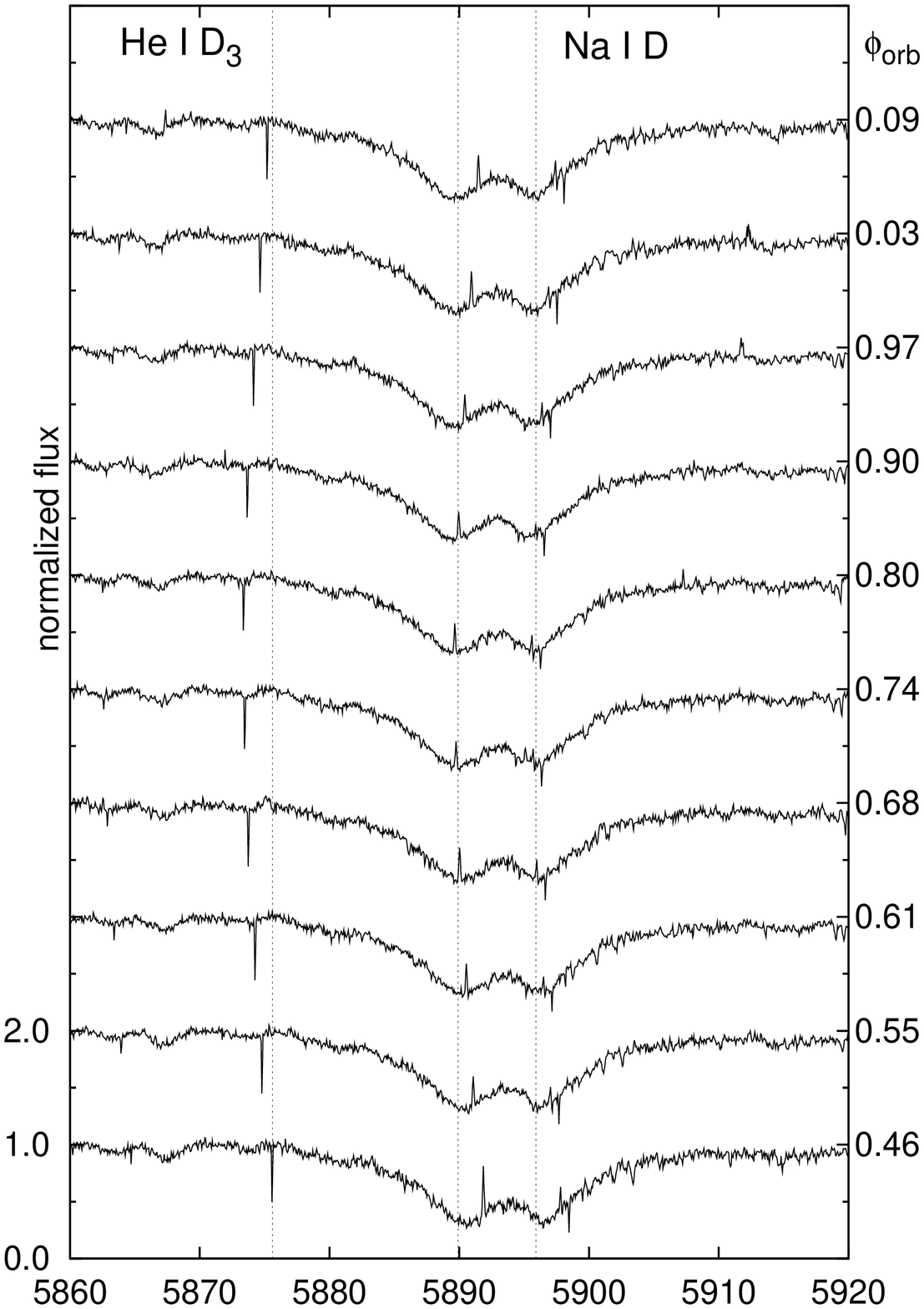}

 \caption{H$\alpha$, H$\beta$ and Na D lines from the quiescent state of V405 And in October 28, 1998. The contribution of the secondary can be well seen around the \ha line. Scales and notation is the same as in Fig.~\ref{fig:fler1994_spect}. }
 \label{fig_online:quiescent_spect}
\end{figure*}
}

During the observations, three noteworthy flare events happened. Two of them were observed with the ELODIE spectrograph: on October 18, 1994 and on October 29, 1998. The third one occurred on February 9, 2007, and was detected in $BV(RI)_C$ photometric bandpasses.

\subsubsection{Spectroscopic events}

The 1994 flare was visible in H$\alpha$, H$\beta$ and He I D$_3$ lines. During the flare, the \ha equivalent widths increased (see Fig.~\ref{fig:eq_width}), indicating stronger activity in the chromosphere. On the dynamic spectrum (Fig.~\ref{fig:halpha}, left panel) as well as on the individual spectra (Fig.\ref{fig:fler1994_spect}, left panel) enhancement in the blue wing of the \ha line is observed since the beginning of the observations.  The flare eruption started at phase 0.51, reached its maximum at phase 0.58 and lasted for more than 2.5 hours, at least until the end of the observations. At the time of the flare H$\beta$ is also increased, and asymmetry is seen in the red wing at all phases (Fig.~\ref{fig:fler1994_spect} middle panel). As the indicator of the explosive event He I D$_3$ appears as prominent emission line (Fig.~\ref{fig:fler1994_spect}, right panel) at phases 0.54--0.58. A very similar event is reported on LO~Peg by \cite{1999A&A...341..527E}.

The weaker flare from October 29, 1998 shows increased \ha emission, and the H$\beta$ level is also higher, but only a marginal He I D$_3$ emission is seen (see Fig.~\ref{fig_online:fler1998_spect} in the online version). This weaker flare was not accompanied by a (large) prominence, the \ha line is quite symmetric, except before and during the flare at phases 0.06--0.14 (Fig.~\ref{fig:halpha}, middle panel). The \ha line outside the small flare is the same as in the quiescent \ha state (Fig.~\ref{fig:halpha}, right panel). The H$\beta$ line is again increased in the red wing at all phases. 

The increase of the blue wing of the  H$\alpha$ line is probably the signature of prominence formation, i.e. emerging material in the magnetic loop with an upflow of $\sim$40km s$^{-1}$ (see Fig.~\ref{fig_online:dynamic} in the online version) which forms a prominence.
After the flare the H$\alpha$ line is again symmetric, since most of the material is heated up and leave the loops during the fast phase of the flare eruption. The red asymmetry of the H$\beta$ line could originate from the downflow of cooling material along the flux loops. 

The H$\alpha$ line of the secondary star is much better visible on the spectra made at the quiescent state of the binary (Fig. \ref{fig_online:quiescent_spect} in the online version), which is partly due to the better signal-to-noise ratio of these spectra, but also the less active primary at this time does not wash out the weaker signal of the secondary. During the observations of the quiescent phase, in the beginning, some red asymmetry is seen again in the H$\beta$ line, indicating that chromospheric flows are present on the star in most of the time.

\subsubsection{Photometric flare}

The light curve of the flare on February 2007 is plotted in Fig. \ref{fig:bigflare}. The system at the peak of the flare is 0.6 magnitudes brighter in $B$ than the quiescent light, and the flare is well seen also in $I_C$ filter, indicating a very strong outburst. The event could be observed for 3.72 hours, which is about one-third of the rotation. Using photometric data, unfortunately no details can be determined for the position of the flare. It is possible that the flare lasted longer than $\sim$4 hours, but the flaring region moved out of view because of the rotation. Since both components are active, the flare could occur in either of the two components.

Assuming that the whole light curve of the flare was observed, we can make an estimation of the flare energies. The energy emitted by the flare in $B, V, R_C$ and $I_C$ bands are 
1.43e+35,
7.58e+34,
1.34e+35 and
1.31e+35 ergs, assuming that the flare occurred on the primary. 

If we suppose that the flare occurred on the much fainter secondary, we need to know its magnitudes for calculating flare energy. According to \chil, the secondary component is at least 3 magnitudes fainter than the primary. If we assume solar metallicity of the system, and a mass of $0.21M_\odot$ from our results for the secondary, using the tables of \cite{2003AJ....126..778V}, we get $B=16.1, V=14.4, R_C=13.1, I_C=11.5$ magnitudes as an upper limit of the secondary component's brightness. Using these values and taking into account the uncertainties in stellar temperatures and radii (see Table \ref{tab:params}) the resulting flare energies satisfactorily agree with the values above, that we determined supposing the flare occured on the primary.
For the details of flare energy  calculating, see e.g. \cite{2007AN....328..904K}. 

After the peak of the flare, four smaller, post-flare eruptions were observed. These oscillations could be the result of reconnection between the moving footpoints of the expanding magnetic loop causing the flare and the intermittent magnetic field \citep[see][]{2007AN....328..904K,2007ApJ...656L.101A} or individual additional flares. However, the quasiperiodicity of the events makes the first scenario more likely \citep[cf.][]{2005A&A...436.1041M}.

\section{Discussion}
\begin{figure}
 \centering
 \includegraphics[width=0.35\textwidth, angle=-90]{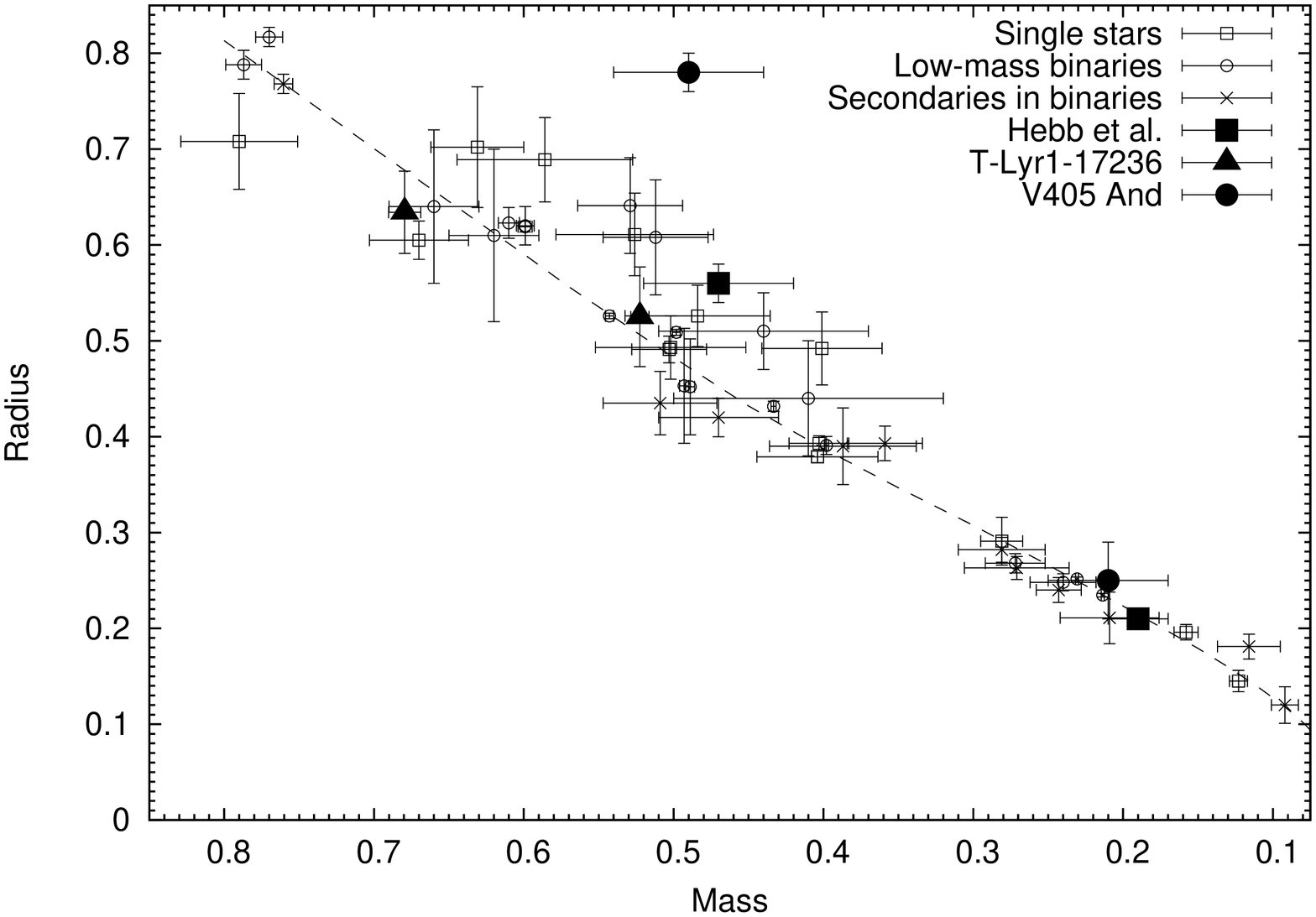}
 % mass-radius.ps: 504x720 pixel, 72dpi, 17.78x25.40 cm, bb=0 0 504 720
\caption{Mass--radius curve for 5 Gyr from \cite{1998A&A...337..403B} (dashed line) and observed values from \cite{2007ApJ...660..732L}, \cite{2008ApJ...684..635B}, \cite{2006astro.ph.10225L} and \cite{2008MNRAS.386..416B}. Three systems, a binary from NGC 1647 \cite{2006AJ....131..555H},  T-Lyr1-17236 \cite{2008ApJ...687.1253D} and V405\,And (present paper) are plotted with special characters.}
\label{fig:mass-radius}
\end{figure}

Our measurements show that V405\,And is a very active RS CVn binary. Strong flare events were observed both in photometry and spectroscopy (see Section \ref{sec:flares}). Photometric data shows that the stellar surface is covered by large spotted regions, which are quite stable during a timescale of the observations ($\sim$650 day), i.e., the maximum spot visibility (spotted light curve minima) is near the same phase. The eruption on September 30, 2007 happened at this position, and probably the one on September 29, 2007 near to that. The large flare on October 18, 1994 also appears at the same phase. These evidences point towards the presence of an active nest that has a very long lifetime, more than 10 years. The phase of the eruptive events is near to the secondary minimum, so the feature is probably locked to the binary orbit. Such long-lived active regions have been observed e.g. on WY Cnc by \cite{2007ARep...51..932K}.

According to \chil, $\log({L_X}/{L_{bol}})$ for V405\,And is $-3.1\pm0.14$ which fits well the rotation activity relations of \cite{2000MNRAS.318.1217J} and \cite{2003A&A...397..147P} 

We compared the masses and radii of V405\,And (Table \ref{tab:params}) with the existing measured values of dwarf stars with masses below 0.8 M$_\odot$ in Fig.~\ref{fig:mass-radius}. The theoretical curve for 5 Gyr from \cite{1998A&A...337..403B} is also plotted using $T_{\mathrm{eff},\odot}=5780$ and $M_\mathrm{bol,\odot}=4.72$. Note that at this part of the theoretical mass-radius diagram there is not much difference between isochrones belonging to different ages; low mass stars evolve slowly.
%$$
%\frac{L}{L_\odot}=\frac{4\pi R^2 \sigma T_\mathrm{eff}}{4\pi R_\odot^2 \sigma T_{\mathrm{eff},\odot}}
%.$$
%For the temperature and bolometric magnitude of the Sun we used .
It is well seen that stars less massive than $\sim$0.32M$_\odot$, which is suggested as limit of full convection, fit well the isochrones. Among these low-mass stars are the secondary components of the binary from NGC~1647 and of V405\,And. However, with increasing mass the correlation gets looser. In Fig. \ref{fig:mass-radius} there are three systems in which the masses of the components are very different: T-Lyr1-17236, the binary from NGC~1647 \citep{2006AJ....131..555H} and V405\,And. T-Lyr1-17236 rotates rather slowly with a period of $P_\mathrm{rot}\simeq8.4d$, whereas the other two systems are very fast rotators (V405\,And: $P_\mathrm{rot}\simeq0.465d$, star from NGC~1647:$P_\mathrm{rot}\simeq0.619d$). The radii of the primary components of the two fast rotating systems are well above the theoretically predicted radius values, especially V405\,And is much larger than expected. On the other hand, both components of the slow rotator T-Lyr1-17236 fit rather well the isochrone. 

\cite{2001ApJ...559..353M} found that the radii of low-mass M dwarfs, concerning stellar structure codes, are larger than expected, and that could be due to their strong magnetic fields, which may push the mass limit of full convection towards lower  masses.
\cite{2006Ap&SS.304...89R} compared theory with observation on masses and radii of low-mass stars (both binaries and single stars). From the high precision results of double lined eclipsing binaries he found that theory predicts smaller radii by about 10\% (or higher temperatures by about 5\%) for stars in the  mass range between 0.4--0.8M$_\odot$. Finally, \cite{2007A&A...472L..17C} carried out evolutionary calculations on low-mass stars taking into account magnetic activity, and demonstrated that magnetic fields  alter the evolution, and that the spot coverage yields larger radii and smaller effective temperature. However, the radius of the primary component of V405\,And is even larger comparing to the results of \cite{2007A&A...472L..17C}. Future modellings with magnetic fields are necessary to map the area of M dwarfs in the mass-radius plane. An interesting question is, what would be the effect of the magnetic interaction between a very active, fully convective star with its close, also active component in a binary system, so that the radius of the more massive component becomes even larger than the recent theory predicts, while at the same time the radius of the fully convective star agrees with the prediction.

At present only two systems are known whose primaries are about 2.5 times more massive than the secondaries, and that one component is above while the other is below the limit of the full convection. The primary of V405\,And deviates most from the expected radius of its mass from the known sample. On the other hand, the secondary of V405\,And would be fully convective even in the presence of its strong magnetic field according to the calculations of \citet[Fig. 1]{2001ApJ...559..353M}. It would be a challenge to model evolution of such binaries whose components are on the two sides of the mass limit of full convection. Of the two known examples, V405\,And is the one with its $V\approx$ 11 mag. brightness which is well observable showing all the signatures of strong magnetic activity like spots and flares. We recommend this binary to the attention of future investigators.

\begin{acknowledgement}
%The financial support  supported by the Hungarian Science Research Program (OTKA) grant T-048961 and T-068626.
Our sincere thanks are due to A. Pr\v{s}a for his suggestions in binary modelling.
An anonymous referee helped to improve the paper considerably.
The financial support of OTKA grant T-068626 is acknowledged.
ZsK is a grantee of the Bolyai J\'anos Scholarship of the Hungarian Academy of Sciences.
This research has made use of the NASA/IPAC Extragalactic Database (NED) which is operated by the Jet Propulsion Laboratory, California Institute of Technology, under contract with the National Aeronautics and Space Administration.
\end{acknowledgement}
\bibliographystyle{aa}
\bibliography{mn-jour,mybib}
\end{document}